\documentclass[aps,prd,eqsecnum,twocolumn,epsf]{revtex4}
\usepackage{amsfonts,latexsym,eucal,amsmath,amssymb,times}
\usepackage[dvips]{graphicx}

\newcommand{\vect}[1]{\!\!\!\mbox{ \boldmath $#1$}}

\begin{document}

\thispagestyle{empty}


\title{Cosmological Density Fluctuations in Stochastic
Gravity \\
-- Formalism and Linear Analysis --}

\author{Yuko Urakawa$^{1}$}
\email{yuko_at_gravity.phys.waseda.ac.jp}
\author{Kei-ichi Maeda$^{1\,,2\,,3}$}
\email{maeda_at_waseda.jp}
\address{\,\\ \,\\
$^{1}$ Department of Physics, Waseda University,
Okubo 3-4-1, Shinjuku, Tokyo 169-8555, Japan\\
$^{2}$ Advanced Research Institute for Science and Engineering,
 Waseda University,
Okubo 3-4-1, Shinjuku, Tokyo 169-8555, Japan\\
$^{3}$ Waseda Institute for Astrophysics, Waseda University,
Okubo 3-4-1, Shinjuku, Tokyo 169-8555, Japan}



\preprint{200*-**-**, WU-AP/***/**, hep-th/*******}


\begin{abstract}
We study primordial perturbations generated from quantum fluctuations
 of an inflaton based on the formalism of stochastic gravity. 
Integrating out the degree of freedom of the inflaton field, we
analyze the time evolution of the correlation function of
the curvature perturbation at tree level
 and compare it with the prediction made by 
the gauge-invariant linear perturbation theory. 
We find that our result coincides with
that of the gauge-invariant perturbation theory
if the e-folding from the horizon crossing time
is smaller than some critical value ($\sim |$
slow-roll parameter $|^{-1}$), which is the case for the 
scales of the observed cosmological structures.
 However, in the limit of the superhorizon scale,
we find a discrepancy in 
the curvature perturbation,
which suggests that we should include
the longitudinal part of
the gravitational field in the quantization of 
a scalar field even in stochastic gravity.
\end{abstract}


\maketitle


\section{Introduction}
Inflation has become the leading paradigm of the early universe not only
because it solves the theoretical difficulties of the standard 
Big-Bang scenario but also because it explains the origin of the
almost scale-invariant primordial density perturbations
 which have been found by
the observation of the cosmic microwave
background radiation (CMB). 
However, we may still not know some fundamental 
parts of the inflationary scenarios or models,
mainly due to our ignorance of physics on the very short scale. It follows that we are largely interested in 
theoretical predictions about what we can learn about inflation from
CMB~\cite{Lidsey:1995np,Bassett:2005xm,Lyth:2007qh,Linde:2007fr}. 
 
In order to analyze some inflation model by the observational data
 of CMB, it is necessary to evaluate the primordial
perturbations generated during inflation. 
We believe that quantum fluctuation of an inflaton field 
gives the origin of seeds of cosmic large-scale structures.
So it may be important to reveal how such a quantum fluctuation
in the inflationary phase
becomes classical density perturbations.
So far there have been a few
 approaches~\cite{Starobinsky:1986fx,Nakao:1988yi,Nambu:1988je,Morikawa:1989xz,Morikawa:1987ci,Tanaka:1997iy,Matarrese:2003ye,Liguori:2004fa, Polarski:1995jg, Lesgourgues:1998gk, Kiefer:1999sj, Kiefer:2006je, Lyth:2006qz}. 
Among them, stochastic gravity may provide
one of the most systematic approaches
\cite{Hu:2003qn,Hu:1989db,Hu:1999mm,Martin:1999ih,Martin:2000dda,Hu:2002jm,Hu:2004gf}. 
It can describe a transition from quantum fluctuations to classical
perturbations systematically. 
Stochastic gravity has been proposed in
order to describe the behaviour of the gravitational field on the sub-Planck,
scale which is affected by quantum matter fields.
On this energy scale, the quantum effect of the
gravitational field may be ignored compared with
quantum fluctuations of matter fields.
Hence the gravitational field can be treated as a classical one.
The semi-classical approach is justified. 

In stochastic gravity, in order to find an effective action, 
we integrate out only the degree
of matter fields by use of the closed time path (CTP)
\cite{Schwinger:1960qe,Chou:1984es,Jordan:1986ug,Calzetta:1986ey}.
This is one way to perform coarse-graining\cite{Su:1987}.  
As a result, we obtain the 
effective equation of motion for gravitational field under the
influence of quantum fluctuation of matter fields, including those
non-linear quantum effects. The obtained evolution equation is the Langevin
type, which contains a stochastic source and a memory term. 

Stochastic gravity looks similar to stochastic
inflation, which was first proposed by Starobinsky 
\cite{Starobinsky:1986fx,Nakao:1988yi,Nambu:1988je,Morikawa:1989xz,Morikawa:1987ci,Tanaka:1997iy,Matarrese:2003ye,Liguori:2004fa}. 
In stochastic inflation,
in order to discuss the evolution of long wave modes
of an inflaton scalar field, which play a crucial role in 
cosmological structure formation,
the inflaton field is split  into two parts: 
superhorizon modes and subhorizon ones. 
The long wave modes 
are affected by  quantum
fluctuations of the short wave modes.
Then the effective equation of
motion for the in-in expectation values
 of superhorizon modes can be
derived by integrating out only subhorizon modes. 
In most works on stochastic inflation, 
the scalar field is discussed in a homogeneous and isotropic spacetime, 
and superhorizon modes and
subhorizon ones interact with each other through a self-interaction of
the scalar field. 

In the last few years, some parts of
quantum fluctuation of the gravitational field
have been taken into account, by adopting 
a gauge-invariant variable as a canonical one to be quantized.
This gauge-invariant variable includes not
only a scalar field but also a longitudinal part of
the gravitational field
\cite{Rigopoulos:2004gr,Rigopoulos:2004ba,Rigopoulos:2005xx,Rigopoulos:2005ae}. Rigopoulos and Shellard have derived
a non-linear evolution equation for long wave modes of the Sasaki-Mukhanov
variable. The equation incorporates 
full non-linear dynamics on large scale. 
However, the contribution from quantum fluctuations
of the short wave modes is evaluated based on 
the linear perturbation equations.
Then only the effect from 
tree-level short wave modes
is taken into account. 

In stochastic gravity, on the other hand,
although the way of the
course-graining is different from that in stochastic inflation, 
the evolution equations in stochastic gravity are 
similar, and both equations can describe the transition from quantum
fluctuations to classical perturbations.
Furthermore, stochastic gravity can incorporate
nonlinear quantum effects of a scalar field.
This gives another advantage in stochastic gravity, 
as we will show below.

When we discuss only the leading part of the curvature
perturbation ($\zeta$) by means of linear analysis,
in spite of mutual difference at the level of microphysics, it
implies that most inflationary models may  be compatible with the
observational data. This is because
this gauge invariant variable $\zeta$, which  is directly related to the
temperature fluctuation of CMB, becomes constant in
the superhorizon region
\cite{Wands:2000dp,Malik:2003mv,Lyth:2004gb}. 
In order to make a difference between many inflationary models,
it is necessary to subtract more information from the observable. 
For this purpose,
non-linear effects have been studied intensely
\cite{Bartolo:2004if,Maldacena:2002vr,Seery:2005wm,Seery:2005gb,Weinberg:2005vy,Weinberg:2006ac,Sloth:2006az,Sloth:2006nu,Seery:2007we,Seery:2007wf}. 
Among such approaches to 
non-linear effects, stochastic gravity would be
 well-suited to compute the loop corrections 
induced from interaction between a scalar
field and the gravitational field. 
It is because the CTP effective
action includes also the non-linear effect of quantum fluctuations 
of a scalar field,
and the effective equation makes it possible to discuss the
loop corrections to primordial perturbations. 

Roura and Verdaguer have applied the formalism of stochastic
gravity to analyse the evolution of the primordial density
perturbation~ \cite{Roura:1999qr, Roura:2007jj}. They discuss the
evolution of Bardeen's gauge-invariant variable $\Phi$ under the
 approximation of de Sitter background spacetime. 
They showed that the order of the amplitude for $\Phi$ 
is the same as the prediction by quantization of the gauge-invariant variable.
However, even at the tree level, so far, the time evolution
of the curvature perturbation $\zeta$ in the superhorizon region, which is
important to compare with the CMB observation, has 
not been sufficiently discussed in stochastic gravity. 
Hence, before we will 
discuss the non-linear effects, it may be better to  reformulate
how to evaluate the primordial perturbations in stochastic gravity. 
Since stochastic gravity
includes  additional effects such as a coarse-graining 
procedure to describe
the transition from quantum fluctuations to classical perturbations, 
it is not 
trivial whether stochastic gravity predicts the same evolution of
the primordial perturbations as that of the gauge-invariant
approach. 
In this paper, we will consider the 
evolution of the curvature perturbation in a uniform density slicing.
 We find that stochastic gravity gives the same result
as that in the gauge-invariant approach except 
for a limited case. 
The time evolution is characterized by
the ratio of the Hubble horizon
scale to the physical scale of fluctuation. 
We find that the amplitude of $\zeta$  in
stochastic gravity deviates from the prediction of the gauge-invariant
perturbation theory only when the ratio
becomes nearly equal to zero.
We  discuss why two quantization procedures do not produce the same
result. 
We will discuss loop corrections both to scalar and
tensor perturbations in \cite{YU20072}.

The paper is organized  as follows. In Sec. \ref{gauge
invariant}, we briefly summarize the quantization of gauge-invariant
variables. In Sec. \ref{SSG}, we present the basic idea of stochastic
gravity, and consider  the basic equations in stochastic
gravity, i.e., the Einstein-Langevin equation, which describes the 
time evolution of the gravitational field affected by
a quantum scalar field. 
Then we discuss the perturbations of the Einstein-Langevin equations
around an inflationary background spacetime. 
In Sec. \ref{Correlation function},
we evaluate the correlation function of $\zeta$ and compare the result
with the prediction in the gauge-invariant approach. 
Finally, in Sec.
\ref{Discussions}, we discuss the reason why the prediction in
stochastic gravity does not agree 
 with that in the conventional 
linear theory  in the superhorizon region. 

Throughout in this paper, we consider a single-field inflation,
which Lagrangian is given by 
\begin{eqnarray}
 \mathcal{L} = - \frac{1}{2} \sqrt{-g}[g^{ab} \partial_a \phi
 \partial_b \phi + 2 V(\phi)]
\,,
\end{eqnarray}
where $\phi$ is an inflaton scalar field and $V(\phi)$ is
its potential.
To characterize the slow-roll inflation, 
we use two slow-roll
parameters:
\begin{eqnarray}
\varepsilon \equiv - {\dot{H}\over H^2}  ~~~  {\rm and}
~~~
\eta_V \equiv {V_{\phi \phi}\over \kappa^2 V}
\,,
\end{eqnarray}
where $H=\dot{a}/a$ and 
$\kappa^2 \equiv 8 \pi G$ are the Hubble expansion parameter 
and the reduced gravitational constant, respectively,
and $V_{\phi \phi}\equiv d^2 V/d\phi^2$. 
As time variable, we adopt the
conformal time, $\tau (<0)$, and represent 
the time derivative by a prime.

\section{Gauge-invariant perturbations}  \label{gauge invariant}
In this section, we briefly summarize a conventional approach 
for a quantization of an inflaton field, in which
the gauge-invariant variable is used 
as a canonical variable. 
For more detailed explanations, please refer to
the papers \cite{Mukhanov:1990me,Sasaki:1986hm,Maldacena:2002vr}. 
Throughout this paper,
we follow the notation for perturbed variables 
defined in \cite{Kodama Sasaki}.

In a linear perturbation theory, each momentum mode decouples in
the basic equations. Therefore, it is sufficient to consider 
only one mode with momentum $k$ in the perturbed equations. 
As for the gauge condition, there
are two convenient choices of time slicing to evaluate primordial
perturbations. 
One is the 
slicing such that  the spatial curvature perturbation vanishes,
$\mathcal{R}=0$, which is called the flat slicing. In this
gauge choice, 
the equation of motion for fluctuation of a scalar field,
is described as
\begin{eqnarray}
&& \varphi_f'' + 2 \mathcal{H} \varphi_f' + (k^2 + a^2 V_{\phi \phi})
\varphi_f 
\nonumber \\
&&~~~~
+ [ 2 a^2 V_{\phi} A  
- \phi' (A' + k \sigma_g)]
  = 0,   \label{pKG eq in UC0}
\end{eqnarray}
where $\varphi_f$,
 $A$ and $\sigma_g$ are perturbed variables
of a scalar field, a lapse function and
a  shear of a unit normal vector
 field to a constant time hypersurface, respectively,
and 
$\mathcal{H}\equiv {a'}/{a} = a H$ 
is a reduced Hubble parameter.
Note that  along with fluctuation of a scalar field, 
there appear metric perturbations
 such as $A$ and  $\sigma_g$\cite{Kodama Sasaki}.
In fact, the square bracket term in Eq. (\ref{pKG eq in UC0})
represents 
fluctuation of the gravitational field. 
Using the perturbed Einstein equations
\begin{eqnarray}
 && 2 \Bigl[ \mathcal{H} (A' + k \sigma_g )+2 
 ( \mathcal{H}' + 2 \mathcal{H}^2 ) A 
  \Bigr] = - 2 \kappa^2
  a^2 V_{\phi} \varphi  ~~~~~~~~~~\\
 && \mathcal{H}(A' + k \sigma_g) 
 = -  \kappa^2 \varphi_f [ a^2 V_{\phi}
 + \mathcal{H} \phi' (3- \varepsilon) ] \label{A+ksigmag}
\,,
\end{eqnarray}
Eq. (\ref{pKG eq in UC0}) can be rewritten by means of one
variable, $\varphi_f$, as 
\begin{eqnarray}
 \varphi_f'' + 2 \mathcal{H} \varphi_f' + ( k^2 + a^2 m_{\rm eff}^{~~2})
 \varphi_f = 0 \label{pKG eq in UC}
\,,
\end{eqnarray}
where
\begin{eqnarray}
 m_{\rm eff}^2 \equiv V_{\phi \phi} + \kappa^2 \frac{\phi'}{\mathcal{H}} 
  \Bigl\{ 2 V_{\phi}  + \frac{\mathcal{H} \phi'}{a^2} 
 (3 - \varepsilon) \Bigr\} \label{meff for phif}
\,.
\end{eqnarray}
The contribution from fluctuation of the gravitational field is put
together into the effective mass term.  If we ignore the
contribution from metric perturbations,
together with $V_{\phi \phi}$, which are
suppressed by the slow-roll parameter
$\varepsilon$ and $\eta_V$, 
$\varphi_f$ obeys the evolution equation for a massless field in de
Sitter spacetime, i.e., 
\begin{eqnarray}
 \varphi_f''(\tau) + 2 \mathcal{H} \varphi_f'(\tau) + k^2
  \varphi_f(\tau)  \simeq 0 
\,.
 \label{pKGeq in UC with sr2}
\end{eqnarray}
This is  
why this slicing is preferred in order to consider the evolution
of perturbations in subhorizon region. 
In de Sitter space, we easily find a two-point correlation function for a massless scalar field.
In order to determine 
positive frequency modes, we have to impose initial conditions. 
When we consider only linear perturbations, it is sufficient
to discuss subhorizon modes on an initial time ($\tau_i$), which are
important for structure formation. 
In the subhorizon limit ($|k \tau | \gg 1$), 
it is appropriate to impose that this scalar field behaves
as if it were a free field in Minkowski spacetime, whose mode functions are
given as
\begin{eqnarray}
\varphi_{f, \,\vect{k}} (\tau_i) = \frac{1}{\sqrt{2k}}~e^{-ik \tau_i}
\end{eqnarray}
at the initial time of inflation. 
Then the two-point function 
for $|k \tau |\ll 1$ 
is obtained as
\begin{eqnarray}
 \langle \varphi_{f, \,\vect{k}} (\tau) \varphi_{f, \,\vect{p}} (\tau) \rangle
 &= &(2 \pi)^3 \delta(\,\vect{k} + \,\vect{p}) \frac{H^2}{2 k^3} 
 (1 + k^2 \tau^2)
\nonumber \\
& \simeq &
(2 \pi)^3 \delta(\,\vect{k} + \,\vect{p}) \frac{H^2}{2 k^3}.
\end{eqnarray}
The other convenient gauge choice is
the comoving slicing, on which the energy flux vanishes,
$T^0_{~i}=0$.
It is preferred when we discuss the evolution of perturbations
in the superhorizon region. 
Since the curvature perturbation in this slicing,
$\mathcal{R}_c$, is related to $\varphi_f$ as
\begin{eqnarray}
\mathcal{R}_c = - \frac{H}{\dot{\phi}} \varphi_f 
\,, 
\end{eqnarray}
the evolution
equation for $\mathcal{R}_c$ is obtained from Eq. (\ref{pKG eq in UC}) and
Eq. (\ref{meff for phif}) as
\begin{eqnarray}
\mathcal{R}_{c,\,\vect{k}}''(\tau)
  + 2 \frac{z'}{z}  \mathcal{R}_{c,\,\vect{k}}'(\tau)
  + k^2 \mathcal{R}_{c,\,\vect{k}} (\tau) = 0
\,,
\end{eqnarray}
where
\begin{eqnarray}
z\equiv \frac{a \dot{\phi}}{H} = {\rm sgn}(\dot{\phi}) \frac{a \sqrt{2
  \varepsilon}}{\kappa}
\,.
 \label{eq for Rc}
\end{eqnarray}
Note that this is the exact equation for $\mathcal{R}_c$. To
derive this equation, neither the slow-roll approximation nor the long
wave approximation is imposed. When we consider the large-scale limit
of 
$k \rightarrow 0$, $\mathcal{R}_c$ includes a constant mode function.
Using the approximation
\begin{eqnarray}
\frac{z'}{z} \simeq - \frac{1 + 3 \varepsilon - \eta_V}{\tau}
\,,
\end{eqnarray}
we find 
the equation for $\mathcal{R}_c$  as
\begin{eqnarray}
&&
\mathcal{R}_{c,\,\vect{k}}''(\tau)
  - \frac{2(1 + 3 \varepsilon - \eta_V)}{\tau} 
 \mathcal{R}_{c,\,\vect{k}}'(\tau)
  + k^2 \mathcal{R}_{c,\,\vect{k}} (\tau) = 0
\,.
\nonumber \\
\end{eqnarray}
If we ignore the time evolution of the slow-roll parameters, 
this equation is solved by Hankel functions.
We find general solution as
\begin{eqnarray}
 \mathcal{R}_c = x^{\alpha} \{ c_1 H_{\alpha}^{(1)}(x)  +
 c_2 H_{\alpha}^{(2)}(x) \}, 
 \label{solution of Rc}
\end{eqnarray}
where $x \equiv  - k \tau$ and 
$\alpha = {3}/{2} + 3 \varepsilon - \eta_V $.
$x$ denotes the ratio of the horizon scale to the physical 
size of a perturbation with momentum $k$.
Since both  Hankel functions ($H_{\alpha}^{(1)}(x)$ and
$H_{\alpha}^{(2)}(x)$) behave as $x^{- \alpha}$ when $x\ll 1$, 
we find that
$\mathcal{R}_c$ approaches to  constant  in the superhorizon region.
 Hence, it is sufficient
to evaluate a two-point function of $\mathcal{R}_c$ around the
horizon-crossing time, i.e,
\begin{eqnarray}
\langle \mathcal{R}_c (\,\vect{k}, \eta)
 \mathcal{R}_c (\,\vect{k}', \eta') \rangle
&=& \Bigl( \frac{H_k}{\dot{\phi}_k}  \Bigr)^2
\langle \varphi_f (\,\vect{k}, \eta) \varphi_f (\,\vect{k}', \eta') \rangle
\nonumber \\
&\simeq&
  \frac{(2 \pi)^3}{4 k^3} \delta(\,\vect{k} + \,\vect{k}')
 \frac{(\kappa H_k)^2}{ \varepsilon_k} 
\,,
\nonumber \\
  \label{amp of Rc}
\end{eqnarray}
where the suffix $k$ represents that 
the variables are evaluated at the time when
the fluctuation mode with 
$k$ crosses a horizon. It follows from Eq. (\ref{amp of Rc}) that
the amplitude of the metric perturbation on given comoving scale, $k$, is
determined by the energy density (or Hubble parameter $H$) 
and by deviation of the
equation of state from de Sitter vacuum, which is described by
the slow-roll parameter 
$\varepsilon$ at the time of horizon crossing. Taking into account 
the fact that
the Hubble parameter decreases and the slow-roll
parameter increases as inflation goes on, the description
Eq. (\ref{amp of Rc}) shows that inflation 
provides the red-tilted spectrum, which agrees with
the CMB observation.

\section{Stochastic gravity} \label{SSG}
In this section, first we shortly summarize
the basic points on stochastic gravity. 
The basic equation in stochastic gravity 
describes the
evolution of the gravitational field, whose source term 
is given by quantum matter fields.
The effective action is obtained by integrating
matter fields with  the CTP formalism. We find
the Langevin type equation, i.e, 
the so-called Einstein-Langevin equation
which is analogous to the equation of motion for 
the Brownian particle. In Brownian motion,
the deterministic trajectory is
influenced and modified by stochastic behaviour of the 
environmental source. 
It is worth noting that this Langevin type equation is well-suited not
only to helping us understand the properties of inflation and the origin of 
large-scale structures in the Universe, but also to explaining the 
transition from quantum fluctuations to classical seeds.

The Langevin type equation is characterized by the existence of 
stochastic variables and  memory terms. The stochastic variables
represent quantum fluctuations of the matter fields.
Under the
Gaussian approximation, its statistics is described by the two-point
function $N_{abc'd'}(x_1, x_2)$, which is called a Noise kernel. 
The Noise kernel is determined by
 the imaginary part of the CTP effective action. 
The CTP effective action with a coarse-graining
includes such an imaginary part. 
The effective equation of motion is also
derived from this effective action.
It is usually  a differential-integral equation
representing the non-Markovian nature. 
The integral part, which is
called the memory term, depends on the history of the gravitational
field itself. In fact, its integrand is composed of 
fluctuations of the gravitational field $\delta g_{ab}$ 
and the two-point function $H_{abc'd'}(x_1, x_2)$, 
which is called the Dissipation kernel. 
The Dissipation kernel is determined by the real part of the CTP effective
action. In Sec. \ref{Perturbation of ELeq} and Sec. \ref{Wightman function}, 
we will explicitly evaluate
the Noise kernel and the Dissipation kernel. 
We also analyze the
perturbations of the above Langevin type equation, 
i.e, the Einstein-Langevin 
equation.
In Sec. \ref{Correlation function}, using
the results obtained in this section, 
we will evaluate the correlation
function of the primordial perturbations. 
  
\subsection{Einstein-Langevin equation} \label{Stochastic gravity}
When we consider an interacting quantum system, which includes the
gravitational field on the sub-Planck scale, 
we may expect that 
quantum fluctuation of the matter fields dominates that of the
gravitational field. 
In stochastic gravity, we assume that 
the gravitational field is not quantized but classical
because we are interested in a sub-Planckian scale.
However, it is
important to take into account fluctuation of a gravitational field
which is induced
through interaction with quantum matter fields. 
In order to discuss such dynamics of the gravitational field,
 the CTP formalism is helpful.

The effective action for the in-in expectation value of the
gravitational field is derived by integrating matter
 fields. In \cite{Martin:1999ih}, Martin and 
Verdaguer derived the
effective equation of motion based on the CTP functional technique
applied to a system-environment interaction, more specifically, on the
influence functional formalism by Feynman and Vernon. 
This CTP effective
action contains two specific terms, in addition to
the ordinary Einstein-Hilbert action, describing the induced effects
through interaction with a quantum scalar field. One  is
a memory term, by which the equation of motion depends on the history
of the gravitational field itself. The other is a
stochastic source $\xi_{ab}$, which describes quantum fluctuation of
a scalar field. 
The latter is obtained from the imaginary part of
the effective action, and as such it cannot be interpreted as a standard
action.
Indeed, there appear 
statistically weighted  stochastic noises as a source for the
gravitational field. Under the Gaussian approximation, this stochastic
variable is characterized by the mean value and the
two-point correlation function: 
\begin{eqnarray}
&&
\langle \xi_{ab}(x) \rangle = 0~,
\nonumber \\
&&
 \langle \xi_{ab}(x_1) \xi_{c'd'} (x_2) \rangle = N_{abc'd'}(x_1,~x_2)
\,,   \label{statistics of xi}
\end{eqnarray}
where
the bi-tensor $N_{abc'd'}(x_1,~x_2)$ is the Noise kernel which represents 
quantum fluctuation of the energy-momentum tensor in a background
spacetime, i.e.,
\begin{eqnarray}
&&
 N_{abc'd'} (x_1, x_2) \equiv 
 \frac{1}{4} \mbox{Re} [F_{abc'd'} (x_1, x_2)]
 \nonumber\\
&&
= \frac{1}{8} \langle \{ \hat{T}_{ab}(x_1) - \langle \hat{T}_{ab}(x_1)
 \rangle, \hat{T}_{ab}(x_2) - \langle \hat{T}_{ab}(x_2) \rangle \} 
 \rangle[g]  \label{def of Noise kernel}
\,,
 \nonumber 
\\
\end{eqnarray}
where $\{\hat{X},\hat{Y}\}=\hat{X} \hat{Y}+ \hat{Y}\hat{X}$,
 $g$ is the metric of a background spacetime, and 
the bi-tensor $F_{abc'd'} (x,y)$ is defined by
\begin{eqnarray}
F_{abc'd'} (x_1, x_2)
&\equiv& \langle \hat{T}_{ab}(x_1) \hat{T}_{c'd'}(x_2) \rangle [g]
 \nonumber \\
&&
 - \langle \hat{T}_{ab} (x_1) \rangle [g] ~
 \langle \hat{T}_{c'd'} (x_2) \rangle [g]  \label{def of F}
\,.
\end{eqnarray}
The expectation value for the quantum scalar field is evaluated in the
background spacetime $g$. The Noise kernel and the Dissipation
kernel correspond to the contributions from internal lines or loops of 
the Feynman diagrams,
which consist of propagators of the scalar field
and do not include propagators of the gravitational field as internal
lines. 
Including the above-mentioned stochastic source of 
$\xi_{ab}$, the effective equation of motion for the gravitational
field is written as 
\begin{eqnarray}
&&  G^{ab}[g + \delta g]
 = \kappa^2 \left[
 \langle \hat{T}^{ab} \rangle_R [g + \delta g] + 2 \xi^{ab} 
\right]
\,,
\nonumber \\
 \label{EL eq}
\end{eqnarray}
where $\delta g$ is the metric perturbation induced 
by quantum fluctuation of matter fields 
 and stochastic source
 $\xi_{ab}$ is characterized by the average value and the two-point
 correlation function Eq. (\ref{statistics of xi}). 

Note that this equation is the same as the semiclassical Einstein
equation except for a source term  of stochastic variables $\xi_{ab}$,
which represents quantum fluctuation of energy-momentum tensor of
matter field. 
Furthermore,  the
expectation value of energy-momentum tensor
 includes a non-local effect as follows.
It consists of three terms as
\begin{eqnarray}
&&
\langle \hat{T}^{ab} \rangle_R [g + \delta g]
 = \langle \hat{T}^{ab}(x) \rangle [g]
  + \langle \hat{T}^{(1)ab}[\phi[g], \delta g](x) \rangle [g]
\nonumber \\
&&
~~
  -2  \int d^4y \sqrt{- g(y)} H^{abcd}[g](x,y) \delta g_{cd}(y) +
  O(\delta g^2) \,,
\nonumber \\
\label{Tabf} 
\end{eqnarray}
where $\hat{T}^{(1)ab}$ is the linearized energy momentum tensor and
 $H^{abcd}$ is the  Dissipation kernel. It depends both on the gravitational field and on
a scalar field. The evolution equation for a scalar field 
also depends on the gravitational field.
As a result, the expectation value of the energy-momentum tensor
depends directly on the spacetime geometry and 
indirectly through a scalar field. 
When we perturb a spacetime with metric $g $ to that with
$( g  + \delta g) $, two different changes appear in the
r.h.s. of Eq. (\ref{Tabf}).
 The second term 
represents the direct change, which is expressed in
terms of fluctuation of the gravitational field $\delta g$
as
\begin{eqnarray}
 && \langle \hat{T}^{(1)ab}[\phi[g], \delta g](x) \rangle
 \nonumber \\ && \hspace{0.45cm}
 = \Bigl( \frac{1}{2}g^{ab} \delta g_{cd}
  - \delta^a_{~c} g^{be} \delta g_{de}
  - \delta^b_{~c} g^{ae} \delta g_{de} \Bigr)
 \langle \hat{T}^{cd} \rangle [g]
 \nonumber \\
&& \hspace{0.45cm} 
- \Bigl\{\Bigl(1 - \frac{2~\varepsilon}{3} \Bigr) \rho
 + \frac{\delta \psi^2}{2 a^2} \Bigr\}
 \Bigl( g^{ac} g^{bd} - \frac{1}{2} g^{ab} g^{cd} \Bigr) \delta g_{cd}
\,.
\nonumber \\ \label{expectation of T1}
\end{eqnarray}
where $\delta \psi^2$ is defined by
\begin{eqnarray}
 \delta \psi^2 \equiv \langle \nabla_0 \psi \nabla_0 \psi 
 + \gamma^{ij}  \nabla_i \psi \nabla_j \psi \rangle [g] 
\end{eqnarray}
We can neglect this term safely on the sub-Planck scale because
this term is smaller by the order of $(\kappa H)^2$ than the previous term.
To derive this expression, 
we have used the background evolution equation for a scalar field. 

While the third integral term  in the
r.h.s. of Eq. (\ref{Tabf}) represents the effect from the indirect
change and is characterized by the Dissipation kernel, which is given by
\begin{eqnarray}
 &&
H_{abc'd'}(x_1, x_2) = H^{\rm (S)}_{abc'd'} (x_1, x_2) 
+ H^{\rm (A)}_{abc'd'} (x_1, x_2) 
~~~~~~~~~~ \label{def of Dissipation kernel} \\
 &&
~~~~~~
 H^{\rm (S)}_{abc'd'} (x_1, x_2) 
= \frac{1}{4} \mbox{Im} [S_{abc'd'} (x_1, x_2)]
  \label{Spart of Dissipation kernel}
\\
 &&
~~~~~~
 H^{\rm (A)}_{abc'd'} (x_1, x_2) 
= \frac{1}{4} \mbox{Im} [F_{abc'd'} (x_1, x_2)]
\,,
 \label{Apart of Dissipation kernel}  
\end{eqnarray}
where $S_{abc'd'} (x_1, x_2)$ is defined by
\begin{eqnarray}
 S_{abc'd'} (x_1, x_2)
 \equiv \langle T^* \hat{T}_{ab}(x_1) \hat{T}_{c'd'}(x_2) \rangle [g] 
\,.
\end{eqnarray}
$T^*$ denotes that we take time ordering before we apply the
derivative operators in the energy momentum tensor. As pointed out in
\cite{Martin:1999ih}, only if the background spacetime  
$g$ satisfies the semiclassical Einstein equation, the
gauge invariance of the Einstein-Langevin equation is guaranteed. 
Hence, in this
paper, to guarantee the gauge invariance, we assume the background
spacetime satisfies the semiclassical Einstein
equation.
 
The Einstein-Langevin equation describes 
fluctuation of the gravitational field, which is induced by the
quantum fluctuations of a scalar field. The Einstein-Langevin equation,
 given by Eq. (\ref{EL eq}) includes two
different sources. One is a stochastic source $\xi_{ab}$, whose
correlation function is given by the noise kernel. From
the explicit form of a Noise kernel
(\ref{def of Noise kernel}), we find that 
$\xi_{ab}$ represents the
quantum fluctuation of the energy momentum tensor. The other is an
expectation value of the energy momentum tensor in the perturbed
spacetime $(g + \delta g)$, which includes a memory term.  The integrand
of a memory term consists of a Dissipation kernel and 
fluctuation of the gravitational field. In order to investigate the
evolution for fluctuation of the gravitational field, it is
necessary to evaluate the quantum correction of a scalar field and
calculate the Noise kernel and the Dissipation kernel.  

\subsection{Perturbed Einstein-Langevin equation}
\label{Perturbation of ELeq}
Since the CTP effective action contains the full quantum effect of  
matter fields, it is expected that we can also 
deal with  non-linear quantum
effects such as loop corrections by means of stochastic gravity. Before
we consider such non-linear effects we shall discuss
linear perturbations, 
which correspond to a tree-level effect in the Feynman diagram.
 Note that in a linear perturbation theory in the homogeneous and
isotropic universe, three types of perturbations,  scalar,  vector, and 
tensor perturbations
are decoupled from each other. Then, we  discuss each perturbation
 independently.

We are interested in density perturbations as seeds for structure formation.
 Then, it is sufficient to consider only scalar
perturbations. In this paper,  we choose the metric as
follows: 
\begin{eqnarray}
 ds^2 &=& - a^2(\tau) (1 + 2 \mathcal{A} Y) d \tau^2
- 2 a^2(\tau) \frac{k}{\mathcal{H}}\Phi Y_j d\tau d x^j
\nonumber \\
&& \hspace{0.5cm}
 + a^2(\tau) \gamma_{ij} dx^i dx^j
\,,
\end{eqnarray}
where $a$ and $\gamma_{ij}$ are a scale factor and the 3 metric of 
maximally symmetric space, and $\mathcal{A}$ and $(k/\mathcal{H}) \Phi$
are the lapse function and the shift vector, respectively. The scalar
perturbations can be expanded by a complete set of harmonic function
$Y ( \vect{x})$ on the three-dimensional space and $Y_j$ is defined by
$Y_j \equiv - k^{-1} Y_{|j}$.
In this gauge choice, the spatial curvature perturbation vanishes. 
We will discuss the evolution of the following variable:
\begin{eqnarray}
 \zeta = \frac{1}{2 \varepsilon} \delta_f,  \label{relation bt zdf}
\end{eqnarray}
where 
$\delta_f\equiv \delta\rho/\rho$ is 
 density perturbation.
This variable $\zeta$ is gauge-invariant,
and  turns out to be a curvature perturbation in
a uniform density slicing. 
In the classical perturbation theory, the
energy conservation law implies that this
variable is conserved
in the superhorizon region for a single-field inflation
\cite{Wands:2000dp,Malik:2003mv,Lyth:2004gb}.
$\zeta$ is directly related
to a gravitational potential at the late stage of the universe
 and in
turn to the observed CMB fluctuations.

The density perturbation in the present slicing is given by 
\begin{eqnarray}
 \delta {T}^0_{~0} &\equiv & - \rho \delta_f Y
\nonumber \\
&=&  \delta g_{0c} \langle \hat{T}^{0c}(x) \rangle [g]
 + g_{0c} \{ \langle \hat{T}^{(1)0c}[\phi[g], \delta g](x) \rangle
\nonumber \\
&&   -2  \int d^4y \sqrt{- g(y)} H^{0cde}[g](x,y) \delta g_{de}(y)
 + 2 \xi^{0c} \} \label{relation for df}
\,.
\nonumber \\
\label{density_perturbation}
\end{eqnarray}
Since the
background energy-momentum tensor is given by
\begin{eqnarray}
 \langle \hat{T}^{0a}(x) \rangle [g]
 = g^{0b} \langle \hat{T}^{a}_{~~b} (x) \rangle [g]  = - \rho~g^{0a}
\,,
\end{eqnarray}
the direct contribution from a change of the gravitational
field is described as
\begin{eqnarray}
  \langle \hat{T}^{(1)00}[\phi[g], \delta g](x) \rangle 
 = - \frac{2}{a^2} \Bigl\{ 1 + \frac{\varepsilon}{3} + O((\kappa H)^2) \Bigr\}
 \rho \mathcal{A} Y 
\,.
\nonumber \\
\label{direct_contribution}
\end{eqnarray}
With these two relations (\ref{density_perturbation})  
and (\ref{direct_contribution}), the density perturbation in flat
slicing is written as
\begin{eqnarray}
 \delta_f \simeq - \frac{2 \varepsilon}{3} \mathcal{A}
 + \frac{2}{\rho} ( \delta \rho_m + \delta \rho_{\xi} ),
\label{delta_f}
\end{eqnarray}
where we have defined the density perturbations of the stochastic source
$\xi^a_{~b}$ and of the memory term as follows:
\begin{eqnarray}
 \delta \rho_m 
 &\equiv& \int d^3 \,\vect{x} e^{-i \,\vect{k} \cdot \,\vect{x}}~
 \Bigl[ g^{00} \int d^4y \sqrt{- g(y)} 
\nonumber \\
&&~~~~~~~ 
 \times H_{00c'd'}[g](x,y)
g^{c'e'} g^{d'f'} \delta g_{e'f'}(y) \Bigr] \nonumber \\
  \delta \rho_{\xi} 
&\equiv&
 \int d^3 \,\vect{x} e^{-i \,\vect{k} \cdot \,\vect{x}}~
  \Bigl[- g_{00} ~ \xi^{00}(x) \Bigr].  
\end{eqnarray}
Since the contribution from the memory term, $\delta \rho_m$ , 
contains
the past history of metric perturbations, 
it seems very difficult to deal with such a term. 
However, as we will show in Sec. \ref{Memory term}, 
we can evaluate the memory term, 
rewriting the integral equation into
a differential equation.

The Hamiltonian constraint equation gives a relation between the
gauge-invariant variable $\mathcal{A}$ and 
the density perturbation
$\delta_f$ as
\begin{eqnarray}
 \mathcal{A} =
 \frac{1}{3} \Bigl( \frac{k}{\mathcal{H}} \Bigr)^2 \Phi 
 - \frac{\delta_f}{2}
\,. \label{Hamiltonian constraint}
\end{eqnarray}
Using it, we eliminate $\mathcal{A}$ in Eq. (\ref{delta_f}),
and find
\begin{eqnarray}
 \Bigl(1 - \frac{\varepsilon}{3} \Bigr) \delta_f = 
 \frac{2}{\rho} (\delta \rho_{\xi} + \delta \rho_m )
 + O \Bigl( (k/ \mathcal{H})^2 \Bigr).
\end{eqnarray}
Hence, in the superhorizon region, the two-point function for $\delta_f$ is
expressed in terms of four correlation functions of
$\delta \rho_{\xi}$ and $\delta \rho_m$, i.e,
\begin{eqnarray}
 && \langle \delta_{f \,\vect{k}}(\tau)
  \delta_{f \,\vect{p}}(\tau) \rangle
 \simeq \frac{4}{V(\tau)^2}  
  \Big[~\langle \delta \rho_{\xi \,\vect{k}}(\tau)
 \delta \rho_{\xi \,\vect{p}}(\tau) \rangle
 \nonumber \\ &&
 \hspace{0.5cm}
 +  \langle \delta \rho_{m \,\vect{k}}(\tau)
 \delta \rho_{\xi \,\vect{p}}(\tau) \rangle
 + \langle \delta \rho_{\xi \,\vect{k}}(\tau)
 \delta \rho_{m \,\vect{p}}(\tau) \rangle
  \nonumber \\ &&
 \hspace{0.5cm}
+ \langle \delta \rho_{m \,\vect{k}}(\tau)
 \delta \rho_{m \,\vect{p}}(\tau) \rangle~\Big].
 \label{correlation of df} 
\end{eqnarray}
Here we have used the relation
\begin{eqnarray}
V(\tau) = \Bigl(1 - \frac{\varepsilon}{3}\Bigr) \rho +
O(\rho~(\kappa H)^2)
\,.
\end{eqnarray}
 The four correlation functions in Eq. (\ref{correlation of df}) are
described by the noise and the Dissipation kernels.
We present the details in  Appendix \ref{ND kernel}.
The correlation function for
$\delta \rho_{\xi}$ is given from the Noise kernel as follows:
\begin{eqnarray}
 &&
\langle \delta \rho_{\xi \,\vect{k}}(\tau_1)
  \delta \rho_{\xi \,\vect{p}}(\tau_2)  \rangle 
  \nonumber \\ && ~~~
 \simeq \frac{(2 \pi)^3}{2}~ \kappa^2 \varepsilon V_1 V_2 
\delta (\,\vect{k} + \,\vect{p})
~{\rm Re} \Bigl[ G^+_{~k} (\tau_1,~\tau_2) \Bigr]
\,,
~~~~~~~
\label{cor for rhoxi with G}
\end{eqnarray}
where $G^+_{~k} (\tau_1,~\tau_2)$ is the Wightman function in momentum
space, which will be given by the mode function. We have denoted
$X(\tau_j)$ as $X_j$, where $X(\tau)$ is some function of 
a conformal time $\tau$.

Since the Noise kernel is not a time-ordered correlation function, it
should be described by the Wightman function in momentum space
$G^+_{~k} (\tau_1,~\tau_2)$.  
In order to derive this relation, we have to assume that 
\begin{eqnarray}
 &&
\Big{|} \frac{\dot{\phi}}{a} \frac{\partial}{\partial \tau_1} 
 {\rm Re} \Bigl[ G^+_{~k} (\tau_1,~\tau_2) \Bigr]\Big{|} 
,~\Big{|}  \frac{\dot{\phi}}{a} \frac{\partial}{\partial \tau_2} 
 {\rm Re} \Bigl[ G^+_{~k} (\tau_1,~\tau_2) \Bigr] 
\Big{|} ~~~~~~~~~~ \nonumber 
\\
&&
~~~~~~~~~~~~<<
  \Big{|} \alpha^{(1)} \kappa  V    
 {\rm Re} \Bigl[ G^+_{~k} (\tau_1,~\tau_2) \Bigr]\Big{|}, 
~~~~~~~
\end{eqnarray}
where $\alpha^{(1)} \equiv V_{\phi}/ ( \kappa V)$. The Wightman function
with the initial condition imposed in Sec.\ref{Wightman function} satisfies
this relation in slow-roll inflationary universe. 

While the density perturbation of the
memory term is expressed as follows :
\begin{eqnarray}
&& \delta \rho_{m \,\vect{k}}(\tau)
  \simeq - 2 \varepsilon  \kappa^2 V(\tau)
  \int^{\tau}_{\tau_k} d \tau_1 ~(a_1)^4
\nonumber \\
&&~~\times \left[ \delta \rho_{\xi \,\vect{k}}(\tau_1) +
 \delta \rho_{m \,\vect{k}} (\tau_1) \right]
{\rm Im} \Bigl[ G^+_{~k} (\tau,~\tau_1)
 \Bigr]\,.~~~
   \label{rhom with G}
\end{eqnarray}
As mentioned in Appendix \ref{ND kernel}, since contributions from
subhorizon region will oscillate, we can ignore them in the integral 
of Eq. (\ref{rhom with G}). Hence the
lower bound of the time integral in Eq. (\ref{rhom with G})
may be given by the horizon crossing time,
$\tau_k=-1/k$. 
Since the memory term depends on the past history, the
integrand of the memory term contains $\delta \rho_m$ itself. In
Sec. \ref{Correlation function}, rewriting this form to a differential
equation, we evaluate contributions from the memory term.

Then, once the Wightman function, $G^+_{~k} (\tau,~\tau_2)$, is
determined, we can evaluate the correlation function for $\delta_f$ and
$\zeta$ from Eq. (\ref{correlation of df}), 
(\ref{cor for rhoxi with
G}), and (\ref{rhom with G}). In the next subsection, imposing the
appropriate initial condition, we will find the mode functions, and
in turn the corresponding Green function. 

\subsection{Wightman function} \label{Wightman function}
Here we consider the Wightman function in momentum space,  
\begin{eqnarray}
 G_k^+ (\tau_1,~\tau_2) \equiv
 \psi_{\,\vect{k}}(\tau_1) \psi^*_{\,\vect{k}}(\tau_2) ,
 \label{Wightman ft in k space}
\end{eqnarray}
where $\psi_{\,\vect{k}}(\tau)$ is the mode function of
a quantum scalar field in the inflationary universe. 
It satisfies the wave equation
\begin{eqnarray}
&&
\psi_{\,\vect{k}}\hspace{0.001cm}''(\tau)
 + 2 \mathcal{H} \psi_{\,\vect{k}}\hspace{0.001cm}'(\tau)
 \nonumber \\
 &&
~~~~~ + \{ k^2 + a^2 \kappa^2 V \eta_V  \} \psi_{\,\vect{k}}(\tau) 
= 0
\,.
\label{eq for mode function 0}
\end{eqnarray}
We solve this equation under the slow-roll condition.
Introducing a new variable as
$\tilde{\psi}_{\,\vect{k}}(\tau)\equiv a(\tau)~
 \psi_{\,\vect{k}}(\tau)$,
this equation is rewritten as
\begin{eqnarray}
 \tilde{\psi}_{\,\vect{k}}''(\tau) + [k^2 -  
 \{ 2 - \varepsilon - \eta_V ( 3 - \varepsilon) \}
 \mathcal{H}^2 ] \tilde{\psi}_{\,\vect{k}}(\tau) = 0 , 
~~~~~ \label{eq for the mode function 1}
\end{eqnarray}
where we have used the relation
\begin{eqnarray}
  a^2 \kappa^2 V = a^2 \kappa^2 \rho
 \Bigl(1 - \frac{\varepsilon}{3} \Bigr)
 = 3 \mathcal{H}^2 \Bigl(1 - \frac{\varepsilon}{3} \Bigr) .
\end{eqnarray}

In an inflationary stage on the sub-Planck scale, we can neglect the term
whose magnitude is smaller 
by the order of $(\kappa H)^2$ than that of the leading term.
We also ignore non-linear terms with respect to
the slow-roll parameters.
So  we do not take into account
time evolution of the slow-roll parameters. 
Under these assumptions,
the equation for $\tilde{\psi}$ becomes   
\begin{eqnarray}
 \frac{d^2~}{d x^2} \tilde{\psi} (x) + \Bigl[ 1 - \frac{2 + 3
(\varepsilon - \eta_V )}{x^2} \Bigr] \tilde{\psi} (x) = 0
\,,
\end{eqnarray}
where we have used
 $\mathcal{H} \simeq -1/[(1 - \varepsilon)\tau]$. 
The general solution for this equation is given  by the Hankel functions
as
\begin{eqnarray}
 &&  \tilde{\psi}_{\,\vect{k}}(\tau) = x^{\frac{1}{2}}
 ~\left[~ \tilde{C} H^{(1)}_{\beta} (x) + \tilde{D} H^{(2)}_{\beta} (x)~
\right]\,,
 \label{def of beta}
\end{eqnarray}
where 
$\beta^2 \equiv {9}/{4} + 3(\varepsilon - \eta_V)$,
with  two arbitrary constants $\tilde{C}$ and $\tilde{D}$. It implies 
\begin{eqnarray}
 && \psi_{\,\vect{k}}(\tau)   = \frac{x^{\frac{1}{2}}}{a(\tau)}
 ~\left[~ \tilde{C} H^{(1)}_{\beta} (x) + \tilde{D} H^{(2)}_{\beta} (x)~
\right] .
\end{eqnarray}
We can assume that the mode functions
should have the same form as in Minkowski spacetime, i.e.,
\begin{eqnarray}
\psi_{\,\vect{k}} (\tau_i) = \frac{1}{\sqrt{2k}} 
~e^{- i k
\tau_i}
\,,
\end{eqnarray}
when the wavelength is much shorter than the horizon
scale, i.e., at very early times of the universe.
This fact may be true in the present gauge rather than the comoving gauge.
 Then the mode function and the Wightman function in  momentum
space are given by  
\begin{eqnarray}
 && \psi_{\,\vect{k}} (\tau) =
 \frac{\sqrt{\pi |\tau|}}{~2}~
 \frac{a_i}{a(\tau)}~
 e^{i \frac{(2 \beta + 1)\pi}{4}}~H^{(1)}_{\beta} (x) \\
 && G_k^+ (\tau_1,~\tau_2) 
 = \frac{\pi \sqrt{\tau_1~\tau_2}}{~4}~
 \frac{a_i^2}{a_1 a_2}~
 H^{(1)}_{\beta} (x_1)~H^{(2)}_{\beta} (x_2)\,,
\nonumber \\
\end{eqnarray}
where $x \equiv - k \tau $.
Setting $a_i=1$, the scale factor $a(\tau)$ is given by
$a(\tau) = \Bigl( \frac{\tau_i}{\tau} \Bigr)^{1+ \varepsilon}$. 
Using this expression, the Wightman function is rewritten as
\begin{eqnarray}
 &&
G_k^+ (\tau_1,~\tau_2)
 = \frac{\pi \sqrt{\tau_1~\tau_2}}{~4}
 \Bigl( \frac{\tau_1 \tau_2}{\tau_i^2} \Bigr)^{1 + \varepsilon}
 H^{(1)}_{\beta} (x_1)~H^{(2)}_{\beta} (x_2) 
\nonumber \\
 &&~~=\frac{\pi}{4} \frac{(x_1 x_2)^{\frac{3}{2}}}{k^3}
  \Bigl( \frac{\tau_1 \tau_2}{\tau_i^2} \Bigr)^{\varepsilon}
 (1 - \varepsilon)^2 H_i^2~
 H^{(1)}_{\beta} (x_1)~H^{(2)}_{\beta} (x_2)
\nonumber \\
\label{G+1}
\end{eqnarray}
Here we have used the relation
\begin{eqnarray}
\tau_i^{-2}= (1- \varepsilon)^2 \mathcal{H}_i^2
 = (1- \varepsilon)^2 H_i^2 \,.
\end{eqnarray}
 In order to compute the correlation
functions for $\delta \rho_{\xi}$ and $\delta \rho_m$, it is sufficient to
consider the evolution of the Wightman function in the superhorizon
region. The behaviour of $G_k^+ (\tau_1,~\tau_2)$ in the superhorizon region
is given in Appendix \ref{G in superhorizon}. Substituting the
approximated expression (\ref{Re of G+})
for the real part of the Wightman function
 into Eq. (\ref{cor for rhoxi with G}), we find that the 
correlation for the density perturbation of stochastic variable
$\delta \rho_{\xi}$ is expressed as
\begin{eqnarray}
&&
 \langle \delta \rho_{\xi \,\vect{k}}(\tau_1)
  \delta \rho_{\xi \,\vect{p}}(\tau_2)  \rangle
 \nonumber \\
&& ~~~~~
 \simeq  \frac{(2 \pi)^3}{4}~\delta (\,\vect{k} + \,\vect{p})~
   \varepsilon V_1 V_2 \frac{\kappa^2 H_i^2}{k^3}
    \frac{(x_1 x_2)^{\eta_V}}{x_i^{2 \varepsilon}}
\,.
~~~~~~~~~~
 \label{cor of rhoxirhoxi}
\end{eqnarray}
Similarly, substituting the approximated expression 
(\ref{Im of G+}) for the imaginary part of the
Wightman function  into
Eq. (\ref{rhom with G}), we find that the density perturbation for the memory term is
\begin{eqnarray}
&& \delta \rho_{m \,\vect{k}}(\tau) \simeq
 \frac{\varepsilon V \kappa^2}{3} \frac{x^{\frac{1}{2}}}{a k^2}
 \int^1_x d x_1 (a_1)^3 x_1^{\frac{1}{2}}
 \nonumber \\
&&
 ~~~~~
\times ~\Bigl\{ \Bigl( \frac{x_1}{x} \Bigr)^{\beta} -
 \Bigl( \frac{x}{x_1} \Bigr)^{\beta}  \Bigr\}~
\{ \delta \rho_{\xi \,\vect{k}}(\tau_1) +
 \delta \rho_{m \,\vect{k}} (\tau_1) \}
 \nonumber \\
&&
  \simeq  \varepsilon H^2
 \frac{x^{\frac{3}{2} + \varepsilon} x_i^{2(1 + \varepsilon)}}{k^2}
 \int^1_x d x_1 x_1^{- \frac{5}{2} - 3 \varepsilon}
 \nonumber \\
&&
 ~~~~~
\times  ~\Bigl\{ \Bigl( \frac{x_1}{x} \Bigr)^{\beta} -
 \Bigl( \frac{x}{x_1} \Bigr)^{\beta}  \Bigr\}~
\{ \delta \rho_{\xi \,\vect{k}}(\tau_1) +
 \delta \rho_{m {\,\vect{k}}} (\tau_1) \} .
\nonumber \\
  \label{delta rhom} 
\end{eqnarray} 

\section{Correlation functions in stochastic gravity}
\label{Correlation function} 
As shown in Eq. (\ref{correlation of df}), the
correlation function for the density perturbation in flat slicing 
consists of  four correlation functions :
\begin{eqnarray}
 && f_{\xi \xi} (\tau_1,~\tau_2) \equiv 
  \langle \delta \rho_{\xi \,\vect{k}}(\tau_1)
  \delta \rho_{\xi \,\vect{p}}(\tau_2)  \rangle 
 \label{def of fxixi} \\
 && f_{\xi m} (\tau_1,~\tau_2) \equiv 
  \langle \delta \rho_{\xi \,\vect{k}}(\tau_1)
  \delta \rho_{m \,\vect{p}}(\tau_2)  \rangle
  \label{def of fxim}  \\
 && f_{m \xi} (\tau_1,~\tau_2) \equiv 
  \langle \delta \rho_{m \,\vect{k}}(\tau_1)
  \delta \rho_{\xi \,\vect{p}}(\tau_2)  \rangle
  \label{def of fmxi}  \\
 && f_{m m} (\tau_1,~\tau_2) \equiv 
  \langle \delta \rho_{m \,\vect{k}}(\tau_1)
  \delta \rho_{m \,\vect{p}}(\tau_2)  \rangle
\,.
 \label{def of fmm} 
\end{eqnarray}
Here we have not explicitly shown  the momentum 
dependence in those expressions just for simplicity.
 $f_{\xi \xi}$ has already been given in
Eq. (\ref{cor of rhoxirhoxi}), i.e.,
\begin{eqnarray}
&&
f_{\xi \xi} (\tau_1,~\tau_2)
 \simeq  \frac{(2 \pi)^3}{4}~\delta (\,\vect{k} + \,\vect{p})~
   \varepsilon V_1 V_2 \frac{\kappa^2 H_i^2}{k^3}
    \frac{(x_1 x_2)^{\eta_V}}{x_i^{2 \varepsilon}}
\,.
\nonumber \\
 \label{fxixi} 
\end{eqnarray}
The other three functions contain the contribution from the memory
term. In the next subsection, \ref{Memory term}, 
we present these three correlation functions.
 Then, these functions determine
the correlation functions for $\delta_f$
(or $\zeta$),
which also gives the correlation function
for the curvature perturbation in uniform density slicing. This gauge-
invariant variable $\zeta$ is related to the fluctuation of the
temperature of CMB. 

\subsection{Memory term}  \label{Memory term}
The memory term represents the non-Markovian nature of the effective
equation of motion. The evolution equation depends on
the past history, showing its
non-local nature. The memory term describes interesting phenomena, such
as  dissipation. It plays an important role in  non-equilibrium
systems. Especially, in stochastic gravity, the memory term represents
the indirect dependence on the gravitational field in the energy
momentum tensor. This point has already been emphasised in
Sec. \ref{Stochastic gravity}. 
The analysis, however, bothers
 us due to the
complexity of
the integral equation. Here, rewriting this integral equation into the
differential equation, we evaluate the contribution from the memory term.

The time evolution of the density perturbation for the memory term
$\delta \rho_m$ is described by Eq. (\ref{delta rhom}). It determines 
 the correlation function for $\delta \rho_m$ in
the integral form. The correlation function between $\delta \rho_{\xi}$
and $\delta \rho_m$ satisfies the integral equation:
\begin{eqnarray}
&&
 f_{m \xi} (x,~x_2)  = \varepsilon H^2
 \frac{x^{\frac{3}{2} + \varepsilon} x_i^{2(1 + \varepsilon)}}{k^2}
 \int^1_x d x_1 x_1^{- \frac{5}{2} - 3 \varepsilon}
 \nonumber 
\\
&&
\times
\left[ \Bigl( \frac{x_1}{x} \Bigr)^{\beta} -
 \Bigl( \frac{x}{x_1} \Bigr)^{\beta}  \right]
 \cdot \left[ f_{\xi \xi} (x_1,~x_2) +
 f_{m \xi} (x_1,~x_2) \right] . 
\nonumber \\
\label{eom for fmxi int}
\end{eqnarray}
Taking derivatives of this equation twice, we 
find a differential equation as
\begin{eqnarray}
&&
 \frac{d}{dx} \Bigl\{ x^{-2-2(\varepsilon - \eta_V)}
 \frac{d}{dx} \left[ x^{-2 \varepsilon - \eta_V}
 f_{m \xi}(x, x_2) \Bigr] \right\} 
\nonumber \\
&&~~~~~~
 = 2 \beta \varepsilon  x^{-4 -4 \varepsilon + \eta_V}
 \left[ f_{\xi \xi} (x, x_2) + f_{m \xi}(x, x_2) \right] .
~~~~~~
\end{eqnarray}
We can simplify it to
 the second order
differential equation with a source term, i.e., 
\begin{eqnarray}
&&
 \frac{d^2}{d x^2} f_{m \xi} (x, x_2)
  - \frac{2}{x} \frac{d}{dx} f_{m \xi} (x, x_2)
\nonumber \\
&&~~~~~~
   + \frac{3(\varepsilon + \eta_V)}{x^2} f_{m \xi} (x, x_2)
= 
 \frac{3 \varepsilon}{x^2} f_{\xi \xi} (x, x_2) .
~~~~~~
\label{diff_eq_fmxi}
\end{eqnarray}
Since Eq. (\ref{eom for fmxi int}) is the integral equation, 
we have to impose the initial conditions, i.e.,
\begin{eqnarray}
 f_{m \xi}(1, x_2) 
= \frac{d}{d x} f_{m \xi}(x, x_2)
 \Big|_{x=1} = 0 
\,.
\label{bc for fmxi}
\end{eqnarray}
The time $x=1$ corresponds to the horizon crossing time. 
Using two  independent 
solutions for the homogeneous equations, which are 
 the power-law
functions 
\begin{eqnarray}
 f_1(x) \equiv x^{3 - (\varepsilon + \eta_V)}~~,
 \hspace{.5cm}
 f_2(x) \equiv x^{\varepsilon + \eta_V} 
\,, 
\end{eqnarray}
we find the solution for Eq. (\ref{diff_eq_fmxi})
with initial conditions (\ref{bc for fmxi}) as
\begin{eqnarray}
 &&
 f_{m \xi} (x, x_2) = \varepsilon
 \int^1_x \frac{d x_1}{x_1} 
 \nonumber \\
&&
 \times \Bigl[~
 \Bigl( \frac{x}{x_1} \Bigr)^{\varepsilon + \eta_V}
 - \Bigl( \frac{x}{x_1} \Bigr)^{3 - (\varepsilon + \eta_V)}
   ~\Bigr]
 f_{\xi \xi} (x_1,~x_2) 
\,.
~~~~~ \label{fmxi0}
\end{eqnarray}
Substituting $f_{\xi\xi}$ given in Eq. (\ref{fxixi}) into
Eq. (\ref{fmxi0}), we obtain
 the correlation function for $\delta \rho_{m}$ and
$\delta \rho_{\xi}$ as
\begin{eqnarray}
 f_{m \xi}(\tau_1, \tau_2)
 &=& \langle \delta \rho_{m \,\vect{k}}(\tau_1)
  \delta \rho_{\xi \,\vect{p}}(\tau_2)  \rangle^{(2)}
 \nonumber \\ 
 &\simeq& \frac{(2 \pi)^3}{4}~\delta (\,\vect{k} + \,\vect{p})~
   \varepsilon V_1 V_2 \frac{\kappa^2 H_k^2}{k^3}
   (x_1 x_2)^{\eta_v}
 \nonumber \\
&&
\times
 \Bigl[ (x_1^{- \varepsilon} - 1)
  + \frac{\varepsilon}{3} \{(x_1)^3 - 1\}  \Bigr]  
\,,
 \label{fmxi}
\end{eqnarray}
where we have used the fact
 that the Hubble parameter at the initial time $\tau_i$
is related to that at the horizon crossing time as 
$H_i^2 x_i^{- 2 \varepsilon} \simeq H_k^2$. 
Note that
$f_{\xi m}(\tau_1, \tau_2)=f_{m \xi} (\tau_2, \tau_1)$.

As for 
the correlation function for $\delta \rho_m$, we have
the integral equation,
\begin{eqnarray}
&& f_{m m} (x,~x_2)  
= \varepsilon H^2
 \frac{x^{\frac{3}{2} + \varepsilon} x_i^{2(1 + \varepsilon)}}{k^2}
 \int^1_x d x_1 x_1^{- \frac{5}{2} - 3 \varepsilon}
  \nonumber \\ is obtained
&&~~
\times
~\Bigl[ \Bigl( \frac{x_1}{x} \Bigr)^{\beta} -
 \Bigl( \frac{x}{x_1} \Bigr)^{\beta}  \Bigr]
\cdot
 \left[ f_{\xi m} (x_1,~x_2) +
 f_{m m} (x_1,~x_2) \right]
\,.
\nonumber \\
\end{eqnarray}
It is converted into the equivalent differential equation,
\begin{eqnarray}
&&
   \frac{d^2}{d x^2} f_{m m} (x, x_2)
  - \frac{2}{x} \frac{d}{dx} f_{m m} (x, x_2)
  \nonumber \\
&&~~
  + \frac{3(\varepsilon + \eta_V)}{x^2} f_{m m} (x, x_2) 
\simeq 
 \frac{3 \varepsilon}{x^2} f_{\xi m} (x, x_2)
~~~~~~~
\end{eqnarray}
with the initial  conditions,
\begin{eqnarray}
 f_{m m}(1, x_2) = \frac{d}{d x} f_{m m}(x, x_2)
 \Big|_{x=1} = 0
\,. \label{bc for fmm}
\end{eqnarray}
Using the time evolution of $f_{\xi m} (x_1,~x_2)$,
we determine the time evolution of $f_{m m} (x_1,~x_2)$ as follows:
\begin{eqnarray}
 f_{m m} (x_1, x_2) &=& \varepsilon
 \int^1_{x_1} \frac{d x}{x} \Bigl[~
 \Bigl( \frac{x_1}{x} \Bigr)^{\varepsilon + \eta_V}
 - \Bigl( \frac{x_1}{x} \Bigr)^{3 - (\varepsilon + \eta_V)}
   ~\Bigr]  
\nonumber \\
&&~\times
 f_{\xi m} (x,~x_2)  \nonumber \\
 &\simeq&
   \frac{(2 \pi)^3}{4}~\delta^{(3)} (\,\vect{k} + \,\vect{p})~
   \varepsilon V_1 V_2 \frac{\kappa^2 H_k^2}{k^3}
   (x_1 x_2)^{\eta_V} 
 \nonumber \\
&&~\times
 \Bigl[ (x_1^{- \varepsilon} - 1)
  + \frac{\varepsilon}{3} \{(x_1)^3- 1 \} \Bigr]
\nonumber \\
&&~\times
  \Bigl[ (x_2^{- \varepsilon} - 1)
  + \frac{\varepsilon}{3} \{(x_2)^3- 1 \}  \Bigr] .
  \label{fmm}
\end{eqnarray}
Here, we find the following 
specific property of the memory term. The leading term of the equal
time correlation function, $f_{mm}(\tau, \tau)$, is given by
\begin{eqnarray}
&&
 f_{mm}(\tau, \tau) \simeq
   \frac{(2 \pi)^3}{4}~\delta  (\,\vect{k} + \,\vect{p})~
   \varepsilon V^2 \frac{\kappa^2 H_k^2}{k^3}~
\nonumber \\
&&~~~~~~\times x^{2 \eta_V}~
\Bigl[ (x^{- \varepsilon} - 1) -
  \frac{\varepsilon}{3} \Bigr]^2
\,.
\end{eqnarray}
After this fluctuation crosses the horizon scale, until
$- \varepsilon \log x \approx 1$,
 $x^{- \varepsilon}$ is nearly
equal to unity. Hence the contributions from the memory term are suppressed
by the slow-roll parameter $\varepsilon$, compared to
$f_{\xi \xi}(\tau, \tau)$. In other words, as long as the e-foldings
from the horizon crossing time to time $\tau$, which is given by 
$N_{k \rightarrow \tau} \simeq - \log x$, is smaller than
$1/\varepsilon$, the contribution from the memory term is negligible
in the correlation function for $\delta_f$ (or $\zeta$). 

However, once the e-foldings $N_{k \rightarrow \tau}$ gets larger
than $1/\varepsilon$, the term $x^{- \varepsilon}$ becomes 
much larger, and eventually
 the contribution from the memory term dominates the
direct contribution from the stochastic variable
$f_{\xi\xi}(\tau, \tau)$. In this case, the
correlation function is
approximated as
\begin{eqnarray}
 f_{mm}(\tau, \tau) \simeq
   \frac{(2 \pi)^3}{4}~\delta (\,\vect{k} + \,\vect{p})~
   \varepsilon~ V^2 \frac{\kappa^2 H_k^2}{k^3}~
   x^{2 (\eta_v - \varepsilon)}.
~~~~~
\end{eqnarray}

\subsection{Correlation functions}
Substituting the correlation function for
$\delta \rho_{\xi}$ and $\delta \rho_{m}$, given by 
Eq. (\ref{fxixi}), Eq. (\ref{fmxi}) and Eq. (\ref{fmm})
into Eq. (\ref{fxixi}), we determine the correlation function for the
density perturbation in flat-slicing, $\delta_f$, as
\begin{eqnarray}
 \langle \delta_{f \,\vect{k}}(\tau)
  \delta_{f \,\vect{p}}(\tau) \rangle
  &=& (2 \pi)^3~\delta  (\,\vect{k} + \,\vect{p})~
   \varepsilon  \frac{\kappa^2 H_k^2}{k^3} x^{2 \eta_V}
\nonumber \\
&&~\times
 \Bigl[ x^{- \varepsilon} + \frac{\varepsilon}{3}
  ( x^{3} - 1 ) \Bigr]^2 
\,.
\end{eqnarray}
The density perturbation in flat-slicing, $\delta_f$, is related to the
curvature perturbation
 in uniform density slicing, $\zeta$, by Eq. (\ref{relation bt zdf}).
We  also  find the
correlation function of $\zeta$ as
\begin{eqnarray}
 \langle \zeta_{\,\vect{k}}(\tau) \zeta_{\,\vect{p}}(\tau) \rangle 
 &\simeq& \frac{(2 \pi)^3}{4} \delta  (\,\vect{k} + \,\vect{p})~
  \frac{\kappa^2 H_k^2}{\varepsilon k^3}  
 \nonumber \\
&&~\times x^{2 \eta_V}
\Bigl[ x^{- \varepsilon} + \frac{\varepsilon}{3}
  ( x^{3} - 1 ) \Bigr]^2   \nonumber \\
 &\simeq& \frac{(2 \pi)^3}{4} \delta (\,\vect{k} + \,\vect{p})~
  \frac{\kappa^2 H_k^2}{\varepsilon k^3}~ x^{2 (\eta_V- \epsilon)}
\,.
\nonumber \\
\label{correlation of zeta}
\end{eqnarray}
Note that after the horizon crossing time, until the e-folding
$N_{k \rightarrow \tau}$ approaches to $1/|\eta_V - \varepsilon|$,
the correlation function for $\zeta$ is approximated as
\begin{eqnarray}
 \langle \zeta_{\,\vect{k}}(\tau) \zeta_{\,\vect{p}}(\tau) \rangle 
 \simeq \frac{(2 \pi)^3}{4} \delta  (\,\vect{k} + \,\vect{p})~
  \frac{\kappa^2 H_k^2}{\varepsilon k^3}
\,.
\end{eqnarray}
Then, in this region, the amplitude of the curvature perturbation
$\zeta$ is constant. However, once the e-folding
$N_{k \rightarrow \tau}$ exceeds $1/|\eta_V - \varepsilon|$, it will
deviate from the constant value. 

In order to compare our result with the
prediction from quantization of the gauge-invariant variable summarized in
Sec. \ref{gauge invariant}, it is helpful to show the relation
\begin{eqnarray}
 v - B = - \frac{2}{9(1 + w)} \Bigl( \frac{k}{\mathcal{H}} \Bigr)^2
 \Bigl( \sigma_g - \frac{k}{\mathcal{H}} \zeta \Bigr)
\,,
\end{eqnarray}
which is
derived from the momentum constraint in the uniform density slicing. 
This relation implies that in the superhorizon region, the uniform density
slicing agrees with the comoving slicing\cite{Lyth:2003im}. 
 Hence, the correlation
function for the curvature perturbation $\zeta$
in the gauge invariant
perturbation theory is given by that for
$\mathcal{R}_c$, i.e., Eq. (\ref{amp of Rc}). Comparing
Eq. (\ref{correlation of zeta}) to Eq. (\ref{amp of Rc}), we find that
as long as the e-folding from the horizon crossing time 
 is smaller than $1/|\eta_V - \varepsilon|$, our result
is the same result as that in the gauge-invariant
perturbation theory. 
However, when the e-folding $N_{k \rightarrow \tau}$
becomes larger than $1/|\eta_V - \varepsilon|$, the predictions by
these two approaches are inconsistent. 

In the next section, we discuss the reason why this deviation appears,
and why the amplitude in stochastic
gravity does not predict a constant evolution if the curvature perturbation
is in the the superhorizon region. 

\section{Summary and Discussions} \label{Discussions}
In this paper, we have considered the evolution of the primordial
perturbations in a slow-roll inflationary universe.
Primordial density perturbations, which play
a crucial role at the later structure formation stage,
are generated inside the horizon from the quantum fluctuation of a
scalar field in the beginning of inflation, and then 
stretched out to the superhorizon scale. 

However, since the observable
quantity through the CMB observation is  classical, in order
to constrain an inflation model by the observation, 
it is necessary to consider a transition from quantum
fluctuations to classical perturbations. Coarse-graining is the
well-suited way to understand such a transition. 
Hence we have discussed cosmological perturbations in the inflationary
stage in stochastic gravity.
Based on the naive expectation that
 quantum fluctuation of the gravitational field is
sufficiently small at the
sub-Planck scale, we assume in
stochastic gravity that contributions from
the Feynman diagrams in which the gravitational field propagates as an
internal line are
negligibly small. Then, when we compute the CTP effective action, we
integrate out only the dynamical degree of freedom
 of a scalar field. As a result, from
this coarse-grained effective action, the evolution equation for the
gravitational field is derived, which is called
the Einstein-Langevin equation. 

Although this effective equation makes it possible to discuss 
non-linear quantum effects such as the loop corrections
of a scalar field, in this paper,
 we have first considered the tree-level effect, which is the leading contributor to the primordial
perturbations.
We have derived two-point correlation functions for
density perturbations and the curvature perturbations.
We find that our results are consistent with
those in the gauge-invariant perturbation theory
unless the e-folding from the horizon crossing time
 exceeds some critical
value ($1/|\eta_V - \varepsilon|$).

 However, as pointed out in the previous section,
the curvature perturbation $\zeta$ does not keep constant
in the superhorizon region if the e-folding 
gets beyond the critical value,
 in contrast to the prediction of the
gauge-invariant perturbation theory. 
We expect that such a deviation appears due to
neglect of quantum fluctuation of the gravitational field. 
In stochastic gravity, we
do not integrate out the degree of freedom of quantum fluctuation of the
gravitational field in
the path integral of the CTP effective action.
 In this sense, quantum fluctuation of the
gravitational field is neglected, although we include
the fluctuation of the gravitational field induced by the quantum fluctuations
of matter fields. 
It seems that this missing part 
induces the deviation in the superhorizon region.

In order to understand this point more clearly, it is helpful to
reconsider the time evolution of the mode functions, 
$\psi_{\,\vect{k}}$, which equation is given by 
Eq. (\ref{eq for mode function 0}).
We emphasize that this equation
represents the wave equation for quantum scalar field in the
background inflationary spacetime. 

Comparing Eq. (\ref{eq for mode function 0}) 
with Eq. (\ref{pKG eq in UC0}) for 
fluctuation of a scalar field in flat slicing ($\varphi_f$), 
we find that the equation for $\varphi_f$
contains  the fluctuation of the gravitational field. 
Such a contribution plays a
crucial role in keeping the curvature perturbation constant in
the superhorizon region. In a gauge-invariant theory, 
 the curvature perturbation in comoving slicing, $\mathcal{R}_c$ is related to $\varphi_f$ as 
\begin{eqnarray}
\mathcal{R}_c = - \frac{H}{\dot{\phi}} \varphi_f ,
\end{eqnarray}
and the equation for $\mathcal{R}_c$ is given by
Eq. (\ref{eq for Rc}). 

In comoving slicing, $\mathcal{R}_c$ is the only one dynamical degree of
freedom, and we have a single  
equation for it without neglect of any perturbed variables.
The equation guarantees that the curvature perturbation 
$\mathcal{R}_c$ is constant at a superhorizon scale.
It means that only
when all contributions of perturbations including
fluctuation of the gravitational
field are taken into account,
$\zeta$=const in the superhorizon region is guaranteed. 
If any part of the contributions is ignored, it implies the deviation from the
constant evolution of $\zeta$ in the superhorizon region.

In fact, when we transform $\psi$ into
$\psi_c =  ({H}/{\dot{\phi}}) \psi$, imitating the relation between
$\delta \phi_f$ and $\mathcal{R}_c$, the equation for $\psi_c$
is written as
\begin{eqnarray}
&&
 \psi_{c,\,\vect{k}}''(\tau)
  + 2 \frac{z'}{z}  \psi_{c,\,\vect{k}}'(\tau)
\nonumber \\
&&~
  + \left[k^2 - 2 \mathcal{H}^2 \varepsilon
  \left(3 - \varepsilon - 
 \frac{V_{\phi}}{\dot{\phi} H} \right) \right]
  \psi_{c,\,\vect{k}} (\tau) = 0
~~~~
\end{eqnarray}
Unlike the case of $\mathcal{R}_c$, because of the existence of
the term proportional to
$\varepsilon$ in the coefficient of $\psi_{c,\,\vect{k}}$, 
$\psi_{c,\,\vect{k}}$=constant is no longer the solution 
for the superhorizon scale ($x\ll 1$),
although the deviation is very small in a slow-roll inflation. 

Finally, we consider the reason why neglection of quantum
fluctuation of the gravitational field has influenced the behaviour
of perturbations
in the superhorizon region. 
In this discussion, first we have to distinguish fluctuation of the
longitudinal mode of the gravitational field
induced by matter fields from fluctuation of
gravitons. 

The latter may be very small and
can be ignored at least on the sub-Planckian energy scale,
while the former may 
not be so, because it is difficult to distinguish 
fluctuation of a scalar field from that of the longitudinal part
of the gravitational field. In fact, if we change the gauge condition,
they are mixed up. Hence when we quantize a scalar field, it may be
natural to include the longitudinal gravitational field as 
well ~\footnote{If we assume that the Einstein equation is no longer
valid when we quantize a matter field, i.e, in off-shell state,
we may not need to include the longitudinal modes of gravitational
perturbations.
It is a completely different quantization method, and predicts
the deviation from the result in the gauge-invariant perturbation theory.}.
In the present calculation based on stochastic gravity, 
we ignore both quantum fluctuations of the gravitational field. 
For the sub-Planck scale inflation, however,
even in stochastic gravity,
we should include contributions from fluctuation of
the longitudinal mode which couples to the matter field,
as we discussed above. 
We may have to improve our formulation for 
density perturbations in stochastic gravity.
However, we should emphasize that the deviation in 
the present approach from the gauge-invariant approach
becomes large
only when the e-foldings from the horizon crossing time to the definite
time exceed $1/|\varepsilon - \eta_V|$. 
For example, if we assume that 
$|\varepsilon - \eta_V|$ is around $0.01$, then it implies that this
deviation appears only when the e-foldings from the horizon crossing
time to the end of inflation $N_{k \rightarrow e}$ exceeds about one
hundred. Taking into account that e-folding $N_{k \rightarrow e}$ for
today's Hubble horizon scale is about fifty, we may conclude that
neglect of the longitudinal modes will influence only on a larger
scale than the observed region today.

In stochastic gravity, as we mentioned,
we can calculate the loop corrections, 
because the CTP effective
action includes also the non-linear effect of quantum fluctuations 
of a scalar field,
and the effective equation makes it possible to discuss the
loop corrections to primordial perturbations. 
We will discuss it in \cite{YU20072}.

\acknowledgments

We would like to thank B.L. Hu, A. Roura,
 M. Sasaki, J. Soda, A.A. Starobinsky, T. Tanaka, and  
 E. Verdaguer,
for valuable discussions. 
Y.U. would specially like to acknowledge B.L. Hu, A.A. Starobinsky, 
T. Tanaka, and E. Verdaguer for very fruitful
comments and suggestions.
This work was partially supported 
by the Japan Society for 
Promotion of Science (JSPS) Research Fellowships (Y.U.),
the Grant-in-Aid for Scientific Research
Fund of the JSPS (No.19540308), the
Japan-U.K. Research Cooperative Program,
the Waseda University Grants for Special Research Projects and
 the 21st-Century
COE Program (Holistic Research and Education Center for Physics
Self-Organization Systems) at Waseda University.
We would also like to acknowledge the hospitality of the Gravitation and Cosmology Group at
Barcelona University, where the early stage of the present work was 
done.

\appendix
\begin{widetext}
\vskip .5cm
\section{Noise kernel and Dissipation kernel} \label{ND kernel}
In Appendix \ref{ND kernel}, we present the Noise kernel and the
Dissipation kernel, which are given by the bi-tensors,
 $F_{abc'd'}(x_1, x_2)$ and
$S_{abc'd'}(x_1, x_2)$. 
These expressions are given from integration
of the CTP effective action
in terms of fluctuation of a scalar field. As seen in
\cite{YU20072}, the lowest-order part of these bi-tensors can be
given by 
\begin{eqnarray}
 F_{abc'd'}(x_1,x_2) 
 &=& 
 a_1 a_2 \dot{\phi}_{1} \dot{\phi}_{2}
 ~\{~\delta^0_{~a} \delta^0_{~c'} G^+_{~;bd'}
 + \delta^0_{~a} \delta^0_{d'} G^+_{~;bc'}
 + \delta^0_{~b} \delta^0_{c'}  G^+_{~;ad'}
 + \delta^0_{~b} \delta^0_{d'} G^+_{~;ac'} 
  \nonumber \\[0.1cm]
 &&  +~ \eta_{c'd'}( \delta^0_{~a} G^+_{~;b0'}
 + \delta^0_{~b} G^+_{~;a0'} )
 + \eta_{ab}( \delta^{0}_{~c'} G^+_{~;0d'}
 + \delta^{0}_{~d'} G^+_{~;0c'})
 + \eta_{ab} \eta_{c'd'} G^+_{~;00'}\} 
 \nonumber \\[0.1cm] 
&&
 -~ \eta_{ab} (a_1)^2 ~ \kappa V_1 \alpha_1^{(1)} a_2 \dot{\phi}_2 
 ~ 
 (\delta^{0}_{~c'}~ G_{~;d'} + \delta^{0}_{~d'}~G^+_{~;c'}
 + \eta_{c'd'}~ G^+_{~;0'}) \nonumber \\[0.1cm]
 && 
 -~ \eta_{c'd'} (a_2)^2 ~\kappa V_2 \alpha_2^{(1)} a_1 \dot{\phi}_1 
 ~ 
 (\delta^{0}_{~a}~ G^+_{~;b} + \delta^{0}_{~b}~G^+_{~;a}
 + \eta_{ab}~ G^+_{~;0}) \nonumber \\[0.1cm] 
&& 
   +~ \eta_{ab}^x \eta_{c'd'}^y (a_1 a_2)^2 ~\kappa^2 V_1 V_2
 ~ \alpha^{(1)}_{1} \alpha^{(1)}_{2}  G^+
\label{F2abcd} 
\end{eqnarray}
\begin{eqnarray}
 S_{abc'd'}(x_1, x_2) 
 &=&
 a_1 a_2 \dot{\phi}_{1} \dot{\phi}_{2}
 ~\{~\delta^0_{~a} \delta^0_{~c'} (i G_F)_{;bd'}
 + \delta^0_{~a} \delta^0_{d'} (i G_F)_{;bc'}
 + \delta^0_{~b} \delta^0_{c'}  (i G_F)_{;ad'}
 + \delta^0_{~b} \delta^0_{d'} (i G_F)_{;ac'} \nonumber \\[0.1cm]
 && +~ \eta_{c'd'}( \delta^0_{~a} (i G_F)_{;b0'}
 + \delta^0_{~b} (i G_F)_{;a0'} )
 + \eta_{ab}( \delta^{0}_{~c'} (i G_F)_{;0d'}
 + \delta^{0}_{~d'} (i G_F)_{;0c'})
 + \eta_{ab} \eta_{c'd'} (i G_F)_{;00'}\} 
 \nonumber \\[0.1cm] 
&& 
 -~ \eta_{ab} (a_1)^2 ~ \kappa V_1 \alpha_1^{(1)} a_2 \dot{\phi}_2 
 ~ 
 (\delta^{0}_{~c'}~ (i G_F)_{;d'} + \delta^{0}_{~d'}~(i G_F)_{;c'}
 + \eta_{c'd'}~ (i G_F)_{;0'}) \nonumber \\[0.1cm]
 && 
 -~ \eta_{c'd'} (a_2)^2 ~  \kappa V_2 \alpha_2^{(1)} a_1 \dot{\phi}_1 
 ~ 
 (\delta^{0}_{~a}~ (i G_F)_{;b} + \delta^{0}_{~b}~(i G_F)_{;a}
 + \eta_{ab}~ (i G_F)_{;0}) \nonumber \\[0.1cm] 
&& 
   +~ \eta_{ab}^x \eta_{c'd'}^y (a_1 a_2)^2 ~ \kappa^2 V_1  V_2 
 ~ \alpha^{(1)}_{1} \alpha^{(1)}_{2}  (i G_F)
\,,
\end{eqnarray}
where $\alpha^{(1)} \equiv V_{\phi}/\kappa V \simeq {\rm sgn}(V_{\phi}) 
\sqrt{2\varepsilon}$. There appear the two different Green functions such
as $G^+=G^+(x_1, x_2)$ and $G_F= G_F(x_1, x_2)$. The former is the
Wightman function, and the latter is the Feynman function. To compute the
correlation function $\delta \rho_\xi$ and $\delta \rho_m$, we have to know
$F_{000'0'}(x_1,x_2)$,
$F_{000'i'}(x_1,x_2)$, $S_{000'0'}(x_1,x_2)$,  and
$S_{000'i'}(x_1,x_2)$,
which are written by
\begin{eqnarray}
 && F^{0}_{~~00'0'}(x_1,~x_2)
  = - (a_2)^2 O_{\tau_1} O_{\tau_2}
  G^+ (x_1,~x_2) \label{F20000}  \\
 && F^{0}_{~~00'l'}(x_1,~x_2)
 = - a_2 \dot{\phi}_2 O_{\tau_1} G^+ (x_1,~x_2)_{;l'}
  \label{F2000l} \\
 && S^{0}_{~~00'0'}(x_1,~x_2)
   = - (a_2)^2 
 \Bigl[ \theta(\tau_1 - \tau_2) O_{\tau_1} O_{\tau_2} G^+(x_1,x_2)
 + \theta(\tau_2 - \tau_1) O_{\tau_1} O_{\tau_2} G^+(x_2,x_1) \Bigr]
 \label{S20000} \\
 && S^{0}_{~~00'l'}(x_1,~x_2)
  = - a_2 \dot{\phi}_2 \Bigl[
 \theta(\tau_1 - \tau_2) O_{\tau_1} G^+(x_1,x_2)_{;l'}
 + \theta(\tau_2 - \tau_1) O_{\tau_1} G^+(x_2,x_1)_{;l'}  \Bigr]
 \label{S2000l}
\,,
\end{eqnarray}
where we have defined the operator $O_{\tau}$ as
\begin{eqnarray}
 O_{\tau} \equiv \frac{\dot{\phi}}{a(\tau)} \frac{\partial~}{\partial \tau} 
  + \alpha^{(1)}\kappa V  \label{def of Otau}
\,.
\end{eqnarray} 

Using Eq. (\ref{F20000})-(\ref{S2000l}),
we shall evaluate the noise and Dissipation kernels in order.

\subsection{Noise kernel}
Substituting the expression Eq. (\ref{F20000}) into 
Eq. (\ref{def of Noise kernel}), we  obtain 
\begin{eqnarray}
 \langle \xi^0_{~0}(x_1) \xi^{0'}_{~0'}(x_2) \rangle
 = N^{0~0'}_{~0~~0'} (x_1, x_2)
 = \frac{1}{4} {\rm Re}[ F^{0~0'}_{~0~~0'} (x_1, x_2) ]
 = \frac{1}{4} O_{\tau_1} O_{\tau_2} {\rm Re}[ G^+ (x_1,~x_2) ] .
\end{eqnarray}
Hence we find the correlation function for $\delta \rho_{\xi}$ as
\begin{eqnarray}
 \langle \delta \rho_{\xi \,\vect{k}}(\tau_1)
  \delta \rho_{\xi \,\vect{p}}(\tau_2)  \rangle 
  &=& \int d^3 \,\vect{x}_1  \int d^3 \,\vect{x}_2
  e^{-i \,\vect{k} \cdot \,\vect{x}_1} e^{-i \,\vect{p} \cdot \,\vect{x}_2}
  \langle \xi^0_{~0}(x_1) \xi^{0'}_{~~0'}(x_2) \rangle^{(2)}
  \nonumber \\
  &=& \frac{(2 \pi)^3}{4}~\delta (\,\vect{k} + \,\vect{p})
   O_{\tau_1} O_{\tau_2}\, {\rm Re} \Bigl[ G^+_{~k} (\tau_1,~\tau_2) \Bigr]
  \nonumber \\
  &\simeq& \frac{(2 \pi)^3}{2}~\delta (\,\vect{k} + \,\vect{p})
  \varepsilon \kappa^2 V_1 V_2 
   ~{\rm Re} \Bigl[ G^+_{~k} (\tau_1,~\tau_2) \Bigr] .
 \label{correlation of xixi}
\end{eqnarray}
In the last equality, we have approximated the operator as
$O_{\tau} \simeq \alpha^{(1)} \kappa V $,
because 
the time derivative of the Wightman
function $\partial_{\tau} G^+$ is proportional to the slow-roll
parameter $\eta_V$ (see Appendix \ref{G in superhorizon}). 

\subsection{Dissipation kernel}
Substituting the expression Eq. (\ref{F20000}) and Eq. (\ref{S20000}) into
Eq. (\ref{Apart of Dissipation kernel}) and Eq. (\ref{Spart of Dissipation
kernel}), we obtain 
\begin{eqnarray}
 H^{(2)0}_{~~~~~~00'0'}(x_1,~x_2)
  &=& H^{(2)0}_{A~~~~00'0'}(x_1,~x_2) + H^{(2)0}_{S~~~~00'0'}(x_1,~x_2)
  \nonumber \\
  &=&  \frac{1}{4} {\rm Im} \Bigl[ F^{(2)0}_{~~~~~~00'0'}(x_1,~x_2)
      + S^{(2)0}_{~~~~~~00'0'}(x_1,~x_2) \Bigr]  \nonumber \\
  &=& - \frac{(a_2)^2}{2}~ \theta(\tau_1 - \tau_2) 
    O_{\tau_1} O_{\tau_2} {\rm Im} \Bigl[ G^+(x_1,x_2) \Bigr] .
\end{eqnarray}
Similarly, substituting the expression Eq. (\ref{F2000l}) and
Eq. (\ref{S2000l}) into Eq. (\ref{Apart of Dissipation kernel}) and
Eq. (\ref{Spart of Dissipation kernel}), we find
\begin{eqnarray}
  H^{(2)0}_{~~~~~~00'l'}(x_1,~x_2)
 &=& H^{(2)0}_{A~~~~00'l'}(x_1,~x_2) + H^{(2)0}_{S~~~~00'l'}(x_1,~x_2)
 \nonumber \\
 &=&  \frac{1}{4} {\rm Im} \Bigl[ F^{(2)0}_{~~~~~~00'l'}(x_1,~x_2)
 + S^{(2)0}_{~~~~~~00'l'}(x_1,~x_2) \Bigr] \nonumber \\
 &=& - \frac{a_2}{2}~\dot{\phi}_2~\theta(\tau_1 - \tau_2)
 O_{\tau_1} {\rm Im} \Bigl[ G^+(x_1,x_2)_{;l'} \Bigr] .
\end{eqnarray}
In order to evaluate $\delta \rho_{m,\,\vect{k}}(\tau)$, it is convenient
to introduce the Dissipation kernel in momentum space as
\begin{eqnarray}
 H_{abc'd'}(\tau_1,~\tau_2,~\,\vect{p},~\,\vect{q})
 &=& \delta^{(3)} (\,\vect{p} + \,\vect{q})
~\tilde{H}_{abc'd'}(\tau_1,~\tau_2,~|\,\vect{p}|) \nonumber \\
 &=& \int d^3 \,\vect{x}_1 \int d^3 \,\vect{x}_2~ 
e^{- i \,\vect{p} \cdot \,\vect{x}_1}
 e^{- i\,\vect{q} \cdot\,\vect{x}_2 } H_{abc'd'}(x_1,~x_2) .
\end{eqnarray}
Especially, the components, $(0,0,0',0')$ and $(0,0,0',l')$, are given
by
\begin{eqnarray}
&& \tilde{H}^{(2)0}_{~~~~~~00'0'}(\tau_1,~\tau_2,~k)
 = - \frac{(a_2)^2}{2} (2 \pi)^3 \theta(\tau_1 - \tau_2)
   ~O_{\tau_1} O_{\tau_2} {\rm Im} \Bigl[ G^+_{~k} (\tau_1,~\tau_2)
   \Bigr] 
 \nonumber \\
 && \tilde{H}^{(2)0}_{~~~~~~00'l'}(\tau_1,~\tau_2,~k)
 = -  \frac{a_2}{2}~ \dot{\phi}_2~ (2 \pi)^3 \theta(\tau_1 - \tau_2)
 \frac{k_{l'}}{i}
 ~ O_{\tau_1} {\rm Im} \Bigl[ G^+_{~k} (\tau_1,~\tau_2) \Bigr] .
\end{eqnarray} 
Using this expression, $\delta \rho_{m,\,\vect{k}}(\tau)$ is written as
\begin{eqnarray}
 \delta \rho_{m \,\vect{k}}^{~(2)}(\tau)
 &\simeq& -~2~ \int \frac{d \tau_1}{(2 \pi)^3}~(a_1)^2 
 [~ \tilde{H}^{0~~(2)}_{~~00'0'}(\tau,~\tau_1,~k^a)
 \mathcal{A}_{\,\vect{k}}(\tau_1)  
 + i \frac{k^{l'}}{k} \tilde{H}^{0~~(2)}_{~~00'l'}(\tau,~\tau_1,~k^a)
 \frac{k}{\mathcal{H}} \Phi_{\,\vect{k}}(\tau_1) ~]  \nonumber \\
 &=& \int^{\tau}_{\tau_0} d \tau_1 ~(a_1)^4
 \Bigl[~\frac{1}{3} \Bigl(\frac{k}{\mathcal{H}_1} \Bigr)^2
  \Bigl\{ O_{\tau} + 3 \frac{\dot{\phi}}{a} \mathcal{H}_1 \Bigr\} 
  \Phi_{\,\vect{k}} (\tau_1)
  - \frac{1}{2} O_{\tau} \delta_{f \,\vect{k}}(\tau_1) \Bigr]
  ~ O_{\tau_1} {\rm Im} \Bigl[ G^+_{~k} (\tau,~\tau_1) \Bigr]
 \label{delta rhom0}
\,,
\end{eqnarray}
where we have neglected the contribution from tensor perturbations which
is expected to be much smaller than that from scalar perturbations
($\delta_{f \,\vect{k}}$ and
$\Phi_{\,\vect{k}}$), and for the second equality, we have used the
relation (\ref{Hamiltonian constraint}).

We can neglect the first term in Eq. (\ref{delta rhom0}) as follows. In
superhorizon region ($x_1 < 1$), this term is suppressed by the factor of
$(k/ \mathcal{H}_1)^2$. 
Around horizon crossing time ($x_1 \simeq 1$),
this term is suppressed by the slow-roll
parameter. Neglecting this term and
using the approximation of
$O_{\tau} \simeq \alpha^{(1)} \kappa V$, 
we find the density perturbation
from the memory term $\delta \rho_m$ as
\begin{eqnarray}
 \delta \rho_{m \,\vect{k}}^{~(2)}(\tau)
  &\simeq& - \frac{1}{2} \int^{\tau}_{\tau_k} d \tau_1 ~(a_1)^4
  \delta_{f \,\vect{k}}(\tau_1)~O_{\tau}  
  O_{\tau_1} Im \Bigl[ G^+_{~k} (\tau,~\tau_1) \Bigr]
 \nonumber \\
  &\simeq& - \frac{1}{2} \int^{\tau}_{\tau_k} d \tau_1 ~(a_1)^4
  \delta_{f \,\vect{k}}(\tau_1)~
  2 \varepsilon  \kappa^2  V V_1~ Im \Bigl[ G^+_{~k} (\tau,~\tau_1) \Bigr]
  \nonumber \\
  &\simeq& - 2 \varepsilon \kappa^2  V 
  \int^{\tau}_{\tau_k} d \tau_1 ~(a_1)^4
  \{ \delta \rho_{\xi \,\vect{k}}(\tau_1) +
 \delta \rho_{m \,\vect{k}} (\tau_1) \} ~ {\rm Im} \Bigl[ G^+_{~k} 
 (\tau,~\tau_1) \Bigr] .
 \label{delta rhom2}
\end{eqnarray}
The lower bound of this integration is given by
the horizon crossing time $\tau_k = - 1/k$ because 
in the subhorizon region ($x_1 > 1$),
the Hankel function appearing
 in $G^+_{~k} (\tau,~\tau_1)$ oscillates, 
and then  no accumulative
contribution exists.

\section{Wightman function in the superhorizon region} \label{G in superhorizon}
In this section, we will consider the evolution of the Wightman function
in the superhorizon region, which is given by
\begin{eqnarray}
 G_k^+ (\tau_1,~\tau_2)
 &=& \frac{\pi}{4} \frac{\sqrt{\tau_1 \tau_2}}{a_1 a_2}~
 H^{(1)}_{\beta} (x_1)~H^{(2)}_{\beta} (x_2) .
\end{eqnarray}
Since the Hankel function is expressed in terms of the Bessel function
as follows :
\begin{eqnarray}
 H_{\beta}^{~(1)}(x) = \Bigl\{ H_{\beta}^{~(2)}(x) \Bigr\}^*
 = \frac{-i}{\sin \beta \pi}
 [J_{- \beta}(x) - e^{- \beta \pi i} J_{\beta}(x)]  \label{Hankel by J}
\,.
\end{eqnarray}
The Bessel function is expanded by a power series of $x$ 
in the region of
 $0<x<1$, i.e.,
\begin{eqnarray}
 J_{\beta}(x) = \Bigl( \frac{x}{2} \Bigr)^{\beta} 
 \displaystyle \sum_{m=0}^{\infty}~
 \frac{(-1)^m (x/2)^{2m}}{m!~\Gamma(\beta + m +1)} 
\,.
 \label{Bessel ft} 
\end{eqnarray}
Then we obtain the asymptotic behaviour of the Hankel function in
the superhorizon region as 
\begin{eqnarray}
 && {\rm Re}[ H_{\beta}^{~(1)}(x_1) H_{\beta}^{~(2)}(x_2) ] 
 \simeq 
 \frac{1}{ \Bigl[\sin \beta \pi ~ \Gamma(1- \beta ) \Bigr]^2}
 \Bigl(\frac{x_1x_2}{4} \Bigr)^{- \beta}  \nonumber \\
 && {\rm Im}[ H_{\beta}^{~(1)}(x_1) H_{\beta}^{~(2)}(x_2) ] 
 \simeq
 \frac{\cos (\beta- 3/2)\pi }{\sin^2 \beta \pi}
 \frac{1}{\Gamma(1- \beta) \Gamma(1 + \beta)}~
 \Bigl[ \Bigl( \frac{x_2}{x_1} \Bigr)^{\beta} -
 \Bigl( \frac{x_1}{x_2} \Bigr)^{\beta}  \Bigr]  
\,.
\end{eqnarray}
Using these formulas, the asymptotic behaviour of the Wightman function
in the superhorizon region is given by
\begin{eqnarray}
 && {\rm Re}[ G^{+}_k (\tau_1,~\tau_2)] \simeq \frac{M_r}{(1- \varepsilon)^2}
  \frac{\sqrt{\tau_1 \tau_2}}{a_1 a_2}~
  (x_1 x_2)^{- \beta}
  \simeq M_r \frac{H_i^2}{k^3} (x_1 x_2)^{\frac{3}{2}}
  \Bigl(\frac{\tau_1 \tau_2}{\tau_i^2} \Bigr)^{\varepsilon}
  (x_1 x_2)^{- \beta} \label{Re of G+}  \\
 && {\rm Im}[ G^{+}_k (\tau_1,~\tau_2)] \simeq M_i
  \frac{\sqrt{\tau_1 \tau_2}}{a_1 a_2}~
  \Bigl[ \Bigl( \frac{x_2}{x_1} \Bigr)^{\beta} -
 \Bigl( \frac{x_1}{x_2} \Bigr)^{\beta}  \Bigr] 
\,, \label{Im of G+}
\end{eqnarray}
where the coefficients $M_r$ and $M_i$ are defined by
\begin{eqnarray}
 && M_r \equiv \frac{\pi}{4} (1 - \varepsilon)^2~
 \frac{2^{2 \beta}}{[\sin \beta \pi \Gamma(1 - \beta )]^2}
 = \frac{1}{2} + O(\varepsilon,~\eta_V) \\
 && M_i \equiv \frac{\pi}{4} ~ \frac{\cos (\beta- 3/2)\pi }{\sin^2 \beta \pi}
 \frac{1}{\Gamma(1- \beta) \Gamma(1 + \beta)} =
 - \frac{1}{6} + O(\varepsilon,~\eta_V) .
\end{eqnarray}

In order to evaluate the correlation function of 
$\delta \rho_{\xi}$, it is also necessary to compute the time derivatives
 of the real part of the Wightman function, which are given by 
\begin{eqnarray}
 && \partial_{\tau_1} {\rm Re} [G^{+}_k (\tau_1,~\tau_2)] \simeq 
 \frac{\eta_V}{\tau_1} G^{+}_k (\tau_1,~\tau_2) 
\,,
 \label{D1G+}  \\
 && \partial_{\tau_2} {\rm Re} [G^{+}_k (\tau_1,~\tau_2)] \simeq 
 \frac{\eta_V}{\tau_2} G^{+}_k (\tau_1,~\tau_2) 
\,,
 \label{D2G+}  \\
 && \partial_{\tau_1} \partial_{\tau_2} {\rm Re}[G^{+}_k (\tau_1,~\tau_2)] \simeq 
 \frac{\eta_V^2}{\tau_1 \tau_2} G^{+}_k (\tau_1,~\tau_2) 
\,.
 \label{D1D2G+}  
\end{eqnarray}

\end{widetext}

\bibliographystyle{apsrev} 
\bibliography{sample.bib} 

%


\end{document}